\documentclass[useAMS, usenatbib, usegraphicx]{mn2e}
\usepackage{amssymb}
\usepackage{booktabs}
\newcommand\ApJ{ApJ}
\newcommand\MNRAS{MNRAS}

\newcommand\ApJS{ApJS}
\newcommand\AnA{A\&A}

\def\alf{Alfv\'en\,}
\def\alfc{Alfv\'enic\,}
\def\as{\'a}
\def\es{\'e}
\def\isp{\'i}
\def\bq{\begin{equation}}
\def\eq{\end{equation}}
\def\ee #1 {\times 10^{#1}}
\def\ut #1 #2 { \, \rmn{#1}^{#2}}
\def\u{\bmath{u}}

\def\persec {\, \hbox{s}^{-1}}

\let\grad=\nabla
\newcommand\cross{\bmath{\times}}

\newcommand\betaH{\tilde{\eta}_\mathrm{H}}
\newcommand\betaP{\tilde{\eta}_\mathrm{P}}
\newcommand\etas{\tilde{\eta}_\perp}
\newcommand\bs{\tilde{\sigma}}
\newcommand\bk{\tilde{k}}
\def\curl{{\grad \cross}}
\def\div #1 {\grad \cdot #1}

\def\vp{v_0^\prime}
\def\avp{|\vp|}

\def\ds{\bmath{dS}}

\def\b{\bmath{b}}

\def\v{\bmath{v}}

\def\vB{\bmath{v}_B}
\def\vi{\bmath{v}_i}

\def\ve{\bmath{v}_e}
\def\vi{\bmath{v}_i}

\def\vn{\bmath{v}_n}

\def\J{\bmath{J}}
\def\B{\bmath{B}}
\def\dBx{\delta B_x}
\def\dBy{\delta B_y}

\def\E{\bmath{E}}            

\def\hB{\bmath{b}}

\def\Ep{\bmath{E'}}  
\def\J{\bmath{J}}
\def\Jpa{\bmath{J_\parallel}}  
\def\Jpe{\bmath{J_\perp}}  

\def\dv{\bmath{\delta\v}}

\def\dB{\bmath{\delta\B}}

\newcommand{\delt} [1] {\frac{\partial #1}{\partial t}}

\def\b1{{\bar{\omega}}}
\def\bo2{\bar{\omega}^2}

\title{Hall instability of solar flux tubes in the presence of shear flows}
\author[B.P.Pandey and Mark Wardle]
        {B.P. Pandey and Mark Wardle \\
{Department of Physics \& Astronomy, Macquarie University, Sydney, NSW 2109, Australia} }
\date{\today}
\pagerange{\pageref{firstpage}--\pageref{lastpage}}
\pubyear{2011}
\begin{document}
\maketitle
\label{firstpage} 
\begin{abstract}
\maketitle 
The magnetic network which consists of vertical flux tubes located in intergranular lanes is dominated by Hall drift in the photosphere-–lower chromosphere region ($\lesssim 1\,\mbox{Mm}$). In the internetwork regions with weak magnetic field, Hall drift dominates above $0.25\,\mbox{Mm}$ in the photosphere and below $2.5\,\mbox{Mm}$ in the chromosphere. Although Hall drift does not cause any dissipation in the ambient plasma, it can destabilise the flux tubes and magnetic elements in the presence of azimuthal shear flow. The physical mechanism of this instability is quite simple: the shear flow twists the radial magnetic field and generates azimuthal field; torsional oscillations of the azimuthal field in turn generates the radial field completing feedback loop. The maximum growth rate of Hall instability is proportional to the absolute value of the shear gradient and is dependent on the ambient diffusivity. The diffusivity also determines the most unstable wavelength which is smaller for weaker fields.    

We apply the result of local stability analysis to the network and internetwork magnetic elements and show that the maximum growth rate for kilogauss field occurs around $0.5\,\mbox{Mm}$ and decreases with increasing altitude. However, for $120\, \mbox{G}$ field, the maximum growth rate remains almost constant in the entire photosphere -– lower chromosphere except in a small region closer to the surface. For shear flow gradient $\sim 0.1 \,\mbox{s}^{-1}$, the Hall growth time is about 1 minute near photospheric footpoint. Therefore, network and internetwork regions with intense field in the presence of shear flow are likely to be unstable in the photosphere. The weak field internetwork regions could be unstable in the entire photosphere--chromosphere. Thus the Hall instability can play an important role in generating low frequency turbulence which can heat the chromosphere.
\end{abstract}


\section{Introduction}
Although our understanding of solar magnetism remains incomplete with each observational improvement revealing new magnetic structures, the following canonical picture remains unaltered. The photosphere is threaded by strong vertical magnetic field concentrated in individual elements or flux tubes at intergranular boundaries \citep{SL64}. The field strength within individual magnetic elements are $\gtrsim 1–-2\, \mbox{kG}$ and their typical size is $\sim 100-200\,\mbox{km}$ at the footpoints. Clear spatial correlations between Ca\textsc{II} K line, UV intensity and heating in the chromosphere maps out the underlying photospheric fields quite well \citep{H09, L09}. The magnetic network shows up as a collection of Ca\textsc{II} bright points in the H and K lines and its appearance is taken as a proxy for foot point motion of the flux tubes \citep{H00, H08}. The lifetime of bright points varies between $\sim 200$ to $400\,\mbox{s}$ \citep{KV99}. Isolated intense field concentrations that produce bright points have also been detected in the internetwork region \citep{D09, A10}. Outside the concentration, non—-vertical (possibly isotropic, Almeida \& Gonzalez, 2011) weaker field also exists in the interior of supergranules \citep{L08}. The flux tubes fan out from the photosphere (with filling factor $\lesssim 1 \%$) to chromosphere where individual tubes merge and form a {\it canopy} (with filling factor $\sim 100\%$). 

The search for a viable mechanism that heats the chromosphere has been underway for the past several decades. The original idea of acoustic heating due to photospheric granular motion \citep{B46, S48, S67, CS92, L93, K07} implies that such a mechanism could be important in regions where the magnetic field is dynamically unimportant. Observations by NASA$\textquoteright$s  {\it Transition Region And Coronal Explorer} (TRACE) satellite revealed that there is hardly any acoustic energy flux (about 90 \% deficit)  to heat the nonmagnetic, quiet chromosphere \citep{FC05a, FC05b, FC06}. Subsequent numerical simulations suggested that this huge discrepancy in the energy flux is possibly due to the limited spatial resolution of TRACE \citep{W04, C07}. Recent high-—resolution observations suggest that the total acoustic power is at least twice as large as the largest power observed before \citep{G10}. However, the jury is out on various acoustic heating models of the chromosphere. 

As intense heating in the chromosphere ($\sim 10^7\,\mbox{ergs}\,\mbox{cm}^{-2}\,\mbox{s}^{-1}$) is directly correlated with the strongest magnetic field concentrations in the network and internetwork regions \citep{S89, L09}, the magnetic field must play important role in heating the chromospheric plasma. For example, resistive heating of the plasma due to Pedersen diffusion is one such possibility \citep{G04}. However, recently it has been pointed out that resistive heating may not be dominant in the chromosphere \citep{S05}. Drawing upon the similarity between Earth$\textquoteright$s ionosphere and the Solar atmosphere, the Farley-–Buneman instability, which is possibly responsible for the equatorial electrojet and can be triggered by cross field motion of the partially ionized medium, has been proposed as another plausible heating mechanism of the chromosphere \citep{F08}. It is quite likely that the flow threshold, $\gtrsim 2\,\mbox{km}/ \mbox{s}$, required to drive this electrostatic instability may not be achievable in the chromosphere \citep{G09}. Therefore, the problem of chromospheric heating remains unsolved. 

It is well known that Hall diffusion can not directly heat the plasma and thus has been overlooked as a possible cause of network--internetwork heating. However, Hall diffusion in the presence of velocity shear can destabilise MHD waves \citep{RK05, K08, BGB10, WS11} and thus can indirectly heat the medium by driving turbulence. The crucial elements required to excite Hall instability are (a) the presence of inhomogeneous azimuthal flow, and, (b) Hall drift of the magnetic field. It is quite likely that both these conditions are met in the photosphere—-chromosphere. 

Rotation has been invoked in the past to explain the stability of flux tubes \citep{S84}. Models of spicules also use the concept of rotating flux tubes \citep{KS97}. The swirling macrospicules, manifested through blue and red---shifted emission in the OV line ($629.73\,$${\buildrel_{\circ}\over {\mathrm{A}}}$) at the transition region indicate rotational velocities $\sim 20 -– 30\,\mbox{km} / \mbox{s}$ \citep{PM98}. The recent discovery of convectively driven localised vortex-—type motions (diameters $\lesssim 500\,\mbox{km}$; lifetimes $3-—7\,\mbox{min.}$) which occur in intergranular magnetic concentrations in the photosphere \citep{B08} has raised the possibility that it may play important role in the energy transfer from photosphere to corona. The analysis of SUNRISE observations \citep{B10} suggests the typical lifetime of the vortices is between $5$ and $20$ minutes with a mean $\sim 7.9 \pm 3.2\,\mbox{minutes}$. \cite{W09} analysed the ionized calcium line and found the presence of rotation in the chromosphere which they inferred to be a manifestation of the twisting and braiding of magnetic footpoint in the photosphere. 

Numerical simulation of near--surface solar convection displays turbulent vortex flows at intergranular lanes \citep{Z93, SN98}. The observation of granulation in the very quiet solar region in the disc centre suggests the presence of vortex tubes with typical mean radius $\sim 150\,\mbox{km}$ and rotation period $\sim 30\,\mbox{s}$ considerably shorter than the lifetime of a typical vortex tube ($\sim$ few minutes) \citep{ST10}. Vorticity generation near boundaries of granules has also been seen in numerical simulations of the photosphere \citep{SN98, M10}. Small--scale, intergranular vortices are formed due to interaction of the photospheric plasma with ambient magnetic field \citep{M11, S11}. Recent high resolution simulations of a realistic photosphere shows that the Hall effect generates out—-of—-plane velocity fields with maximum speed $\sim 0.1\,\mbox{km} / \mbox{s}$ at interface layers between weakly magnetized light bridges and neighbouring strong field umbral regions \citep{C12}. To summarise, both observation and numerical simulation point to the frequent occurrence of shear flows in the solar photosphere. Therefore, both of the above mentioned conditions necessary for the onset of Hall instability are likely to be met in the photosphere-chromosphere. 

\begin{figure}
\includegraphics[scale=0.30]{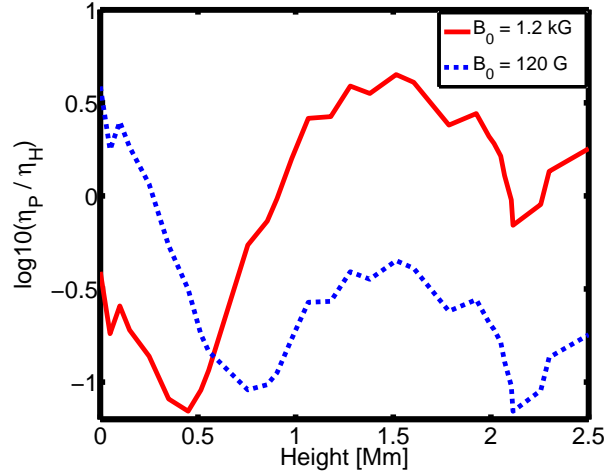}
\caption{The ratio of Pedersen ($\eta_p$) and Hall ($\eta_H$) diffusion is shown in the above figure. The diffusivities have been calculated for magnetic field profile $B = B_0 (n_n / n_0)^{0.3}$ where $B_0 = 1.2\,\mbox{kG}\,,$ (bold line), and $120\,\mbox{G}\,,$ (dotted line), $n_n$ is neutral number density and $n_0$ is  neutral number density at the solar surface, $h = 0$, taken from model C, Vernazza et al. (1981).}
 \label{fig:DF}  
\end{figure}

In Fig.~(\ref{fig:DF}) we show the ratio of Pedersen, and, Hall diffusivities, $\eta_P / \eta_H$. Although both ion and neutral number densities decrease with increasing height, the ratio $B / n_i$ where $n_i$ is ion number density do not always decrease and this is reflected in jagged diffusivity profiles.\footnote{The detailed calculation of various diffusivities for the photosphere–-chromosphere region will be given in \cite{PW12}.}  Fig.~\ref{fig:DF} suggest that Hall dominates Pedersen diffusion below $\sim 1\,\mbox{Mm}$ in the network ($\sim \mbox{kG}$) region. In the internetwork region, where the field is $\sim 100\,\mbox{G}$, Hall diffusion dominates in the entire photosphere above $0.25\,\mbox{Mm}$ and the chromosphere. Thus, conditions (a) and (b) required for the onset of Hall instability are easily met in the solar atmosphere. Clearly, Hall diffusion in the presence of shear flow can destabilise the flux tubes and magnetic elements and drive turbulence in the medium which could facilitate the heating.

In the present paper, we investigate the local stability of the network and internetwork magnetic elements by assuming that magnetic field is immersed in the highly diffusive plasma medium. Since the solar atmosphere is highly stratified, validity of present local analysis is restricted to short (with respect to the scale height) wavelength fluctuations. Further, since flux tube will be approximated by a planar geometry, our analysis is valid only for wavelengths much smaller than the tube radius.         

\section{Basic model}
The partially ionized photosphere--chromosphere plasma consist almost entirely of electrons, protons, singly ionized metallic ions, H, He I, He II, and He III. We shall assume that the plasma consists only of electrons, hydrogen ions and neutrals and neglect the distinction between the metallic and hydrogen ions. The single-fluid description of partially ionized plasma is well known for past several decades \citep{C57, BR65, MK73}. However, only recently \cite{PW08} critically analysed the non—-ideal MHD effects in a partially ionised plasma by developing a unified single-fluid framework for the plasmas of arbitrary ionization. Therefore, we shall utilise single fluid formulation given by \cite{PW08}. Note that plasma quasineutrality is implied in the single fluid formulation.

The continuity equation is 
\bq
\frac{\partial \rho}{\partial t} + \grad\cdot\left(\rho\,\v\right) = 0\,.
\label{eq:cont}
\eq
Here $\rho \approx \rho_i + \rho_n$ is bulk fluid density and $ \v = (\rho_i\,\vi + \rho_n\,\vn)/\rho$ is bulk velocity  with $\rho_i,\,\vi$ and $\rho_n,\,\vn$ representing the mass density and bulk velocities of the ion and neutral fluids respectively. The momentum equation is 
\bq
\rho\,\frac{d\v}{dt}=  - \nabla\,P + \frac{\J\cross\B}{c}\,,
\label{eq:meq}
\eq
where $\J = e\,n_e\,\left(\vi - \ve\right)$ is the current density, $\B$ is the magnetic field and $P = P_e + P_i + P_n$ is the total pressure. 
The induction equation is
\begin{eqnarray}
\delt \B = \curl\left[
\left(\v\cross\B\right) - \frac{4\,\pi\,\eta}{c}\,\J - \frac{4\,\pi\,\eta_H}{c}\,\J\cross\hB
\right. \nonumber\\
\left.
+ \frac{4\,\pi\eta_A}{c}\,
\left(\J\cross\hB\right)\cross\hB
\right]\,,
\label{eq:ind}
\end{eqnarray}
where $\hB = \B /B$ is the unit vector along magnetic field. The expression for Hall, $\eta_H$, ambipolar, $\eta_A$ and, Ohm, $\eta$  diffusivities 
are \citep{PW08}
\bq
\eta_H = \left(\frac{v_A^2}{\omega_H}\right)\,,
\eta_A = D\,\left(\frac{v_A^2}{\nu_{ni}}\right)\,,
\mbox{and}\,,
\eta = \beta_e^{-1}\,\eta_H\,,
\label{eq:ddf}
\eq
where $D = \rho_n / \rho\,,$ $v_A = B / \sqrt{4\,\pi\,\rho}$ is \alf velocity, $\beta_e = \omega_{ce} / \nu_{en}$ is the electron Hall parameter, a ratio of the  electron cyclotron to the electron–-neutral collision frequencies and the Hall frequency $\omega_H$ is
\bq
\omega_H  = \frac{\rho_i}{\rho}\,\omega_{ci} \approx X_e\,\omega_{ci}\,,
\eq  
with $X_e = n_e / n_n$. 

The concept of {\it frozen--in flux} is quite useful in ideal MHD and thus it would be profitable to cast the induction Eq.~(\ref{eq:ind}) in similar form. Such a formulation allows us to visualise the magnetic field as a {\it real physical entity} that is drifting through the fluid with a given velocity. Lets consider magnetic flux  $\Phi  = \int{\B\cdot d\bmath{S}}$ through surface $S$ encircled by an arbitrary closed contour $C$ in the plasma moving with velocity $\u$. The time rate of change of the magnetic flux is given as
\begin{eqnarray}
\frac{d\,\Phi}{dt} = 
\int_S \delt{\B}\cdot\ds + \oint_C \B \cdot \left(\u\cross\bmath{dl}\right)  = \nonumber\\
= - c \oint_C \left( \E + \frac{1}{c}\,\u\cross\B\right)\cdot \bmath{dl}\,
\label{eq:fx1}
\end{eqnarray}
where $\ds$ is the surface element and $\bmath{dl}$ is the line element along $C$. The flux $\Phi$ through the closed contour $C$ is conserved only when the integrand in the last expression of Eq.~(\ref{eq:fx1}) vanishes, i.e. $c\,\E + \u\cross\B = 0$. 

We note that the induction Eq.~(\ref{eq:ind}) is derived from generalized Ohm$\textquoteright$s law 
\bq
\frac{c^2}{4\,\pi}\Ep = \eta\,\Jpa + \eta_H\,\J\cross\hB + \eta_P\,\Jpe\,,
\eq
where electric field is written in the neutral frame and parallel and perpendicular components of the current $\J$ refers to the orientation with respect to the ambient magnetic field
\bq
\Jpa = \left(\J\cdot\hB\right)\,\hB\,,\quad \Jpe = \J - \Jpa\,.
\eq

The Pedersen diffusion $\eta_P$ is related to the ambipolar diffusion as
\bq
\eta_A = \eta_P - \eta\,.
\eq
Thus defining magnetic drift velocity as \citep{WS11}
\bq
\vB = \eta_P\,\frac{\left(\grad\cross\B\right)_{\perp}\cross\hB}{B} -– 
\eta_H\,\frac{\left(\grad\cross\B\right)_{\perp}}{B}\,, 
\label{eq:md0}
\eq
the induction Eq.~(\ref{eq:ind}) can be explicitly written in terms of fluid and field velocities as 
\bq
\delt \B = \curl\left[
\left(\v + \v_B\right)\cross\B - \frac {4\,\pi\,\eta}{c}\,\Jpa\right]\,.
\label{eq:indA}
\eq

We may infer from $c\,\E_{\perp} + \left(\v + \vB \right)\cross\B = 0$ that when each point of the boundary $C$ moves with the velocity 
\bq
\u = \v + \vB\,,
\eq
the time rate of change of flux in partially ionized medium becomes
\bq
\frac{d\Phi}{dt} = \oint_C \,\eta\,\left(\curl\B\right)_{\parallel}\,\cdot\,\bmath{dl}\,.
\label{eq:fx2}
\eq
We see from Eq.~(\ref{eq:fx2}) that the presence of parallel current is responsible for the flux non—-conservation although rate of flux decay is directly proportional to Ohm diffusion. In the absence of parallel current, non—-ideal MHD effects only redistribute magnetic flux in the medium leaving total flux unchanged. At first sight this result appears paradoxical since non—-ideal MHD effects, particularly, Pedersen (Ohm + Ambipolar) diffusion, owing to its dissipative nature, must affect magnetic flux. Indeed energy dissipation is inevitable in such collision dominated weakly ionized plasmas. However, energy loss occurs due to redistribution / relaxation of the magnetic field and not due to flux annihilation \citep{P63}. Therefore, if parallel current is absent in the medium, the total flux is conserved.     

Although for the past two decades, the dynamics of the weakly ionized--weakly magnetized photosphere and weakly ionized--strongly magnetized chromosphere has been investigated in the framework of {\it non-—ideal MHD} [Arber et al. (2007),  Goodman (2011) and references therein] it is pertinent to estimate relative importance of various terms in the induction equation (\ref{eq:ind}). 
\begin{figure}
\includegraphics[scale=0.30]{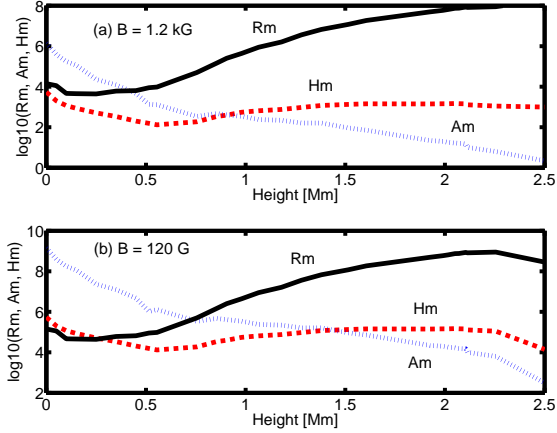}
\caption{The magnetic Reynolds numbers $\mbox{Rm}$, $\mbox{Am}$ and $\mbox{Hm}$ pertaining to Ohm, ambipolar and Hall diffusion is shown against height for  photosphere—-chromosphere plasma parameters of model C, VAL81.}
 \label{fig:DF1}  
\end{figure}
They are on the order of $\sigma\,B : v\,B / L : \eta\,B / L^2 : \eta_H\,B / L^2: \eta_A\,B / L^2\,$. Here we have replaced $\partial_t$ by $\sigma$ which is dynamical frequency set by the signal speed and gradient length scale $L$ \citep{PW08}. The ratio of convective and diffusive terms in the induction Eq.~(\ref{eq:ind}) is known as Magnetic Reynolds number $\mbox{Rm}$ in MHD \citep{GP04}. We shall generalise this definition of magnetic Reynolds number and, introduce three different magnetic Reynolds numbers corresponding to Ohm, ambipolar and Hall diffusion as 
\bq
\mbox{Rm} = \frac{v\,L}{\eta}\,,\quad \mbox{Am} = \frac{v\,L}{\eta_A}\,, \quad \mbox{Hm} = \frac{v\,L}{\eta_H}\,. 
\eq
The ambipolar and Hall Reynolds numbers $\mbox{Am}$ and $\mbox{Hm}$ can be expressed in terms of magnetic Reynolds number $\mbox{Rm}$ with the help of ion and electron Hall parameters. Thus defining ion Hall parameter $\beta_i = \omega_{ci} / \nu_{in}$, where $\omega_{ci} = e\, B / m_i\,c$ is the ion cyclotron frequency, ambipolar and Hall Reynolds numbers can be written as
\bq
\mbox{Am} = \mbox{Rm} / \left(\beta_i\,\beta_e\right)\,,\quad\mbox{Hm} = \mbox{Rm} / \beta_e\,.
\eq  
Clearly, both ambipolar and Hall Reynolds numbers are functions of plasma magnetisation, that is plasma Hall parameter $\beta_e$ and $\beta_i$ \citep{P08} and therefore, their values in weakly ionized and weakly magnetized photosphere and weakly ionized and strongly magnetized chromosphere widely differ. 

In Fig.~(\ref{fig:DF1}) we plot Rm, Hm and Am against height for a kG [Fig.~\ref{fig:DF1}(a)] and $0.1$ kG [Fig.~\ref{fig:DF1}(b)] fields. As has been mentioned earlier, we have utilised model C, \cite{VAL81} (hereafter VAL81) to calculate various diffusivities. Here we assume $L = 100\,\mbox{km}$ and $v = v_A$. We see that for both kG and sub kG fields, magnetic Reynolds number Rm is very large except close to the surface. Therefore, barring the solar surface, Ohm diffusion can be neglected in the entire photosphere--chromosphere. However, unlike Rm, Hall and ambipolar Reynolds number are not very large for a kG field in most of the photosphere—-chromosphere. In fact in the large part of photosphere—-chromosphere ($\gtrsim 0.25\,\mbox{Mm}$), $\mbox{Hm}$ takes  values between $10^2$ and $10^3$. Similarly, $\mbox{Am}$ varies between $1$ and $10^3$ in this interval. Clearly, these quantities are several oreder of magnitude smaller than Rm.  The present estimate of $\mbox{Hm}$ and $\mbox{Am}$ is broadly in agreement with \cite{K11, SK12}. For sub--\alfc motion \citep{HKM10}, both $\mbox{Hm}$ and $\mbox{Am}$ can become of the order of unity or less.  For $0.1\,\mbox{kG}$ field [Fig.~\ref{fig:DF1}(b)], advection term in induction Eq.~(\ref{eq:ind}) is more dominant than Hall and ambipolar diffusion in comparison with the previous case. However, at smaller scales or, sub—-\alf speed, even for a weak field, Hall and ambipolar diffusion becomes important. 

Recent radiative MHD simulation of weakly ionised solar atmosphere shows that Hall term is largest in the lower and middle chromosphere and in the corona. In the intergranular lanes in the photosphere, the Hall term is the most important diffusion term whereas ambipolar diffusion is important in the region from the upper--photosphere to the upper chromosphere \citep{SK12}. In the chromosphere, ambipolar diffusion dominates almost everywhere except in the lower chromosphere in shock fronts. To conclude, both Hall and ambipolar diffusion can compete with the advection term in the induction equation and thus, must be retained while investigating the weakly ionized solar atmosphere.   

\section{Stability Analysis}
As magnetic elements in the solar atmosphere are concentrated into discrete structures, they are often modelled as flux tubes or, flux slabs, implying that the field either posses cylindrical or translational symmetry. Further, both in active and quiet phases, magnetic elements are highly dynamic consisting of numerous flows with different spatial and temporal time scales. Recent two dimensional simulations of umbral magneto convection suggest that the dynamical scale over which Hall effect can generate magnetic and velocity fields  is much faster [$\sim$ 10--20 km spatial scale and $\sim 300\,\mbox{s}$ temporal scale, \cite{C12}] than the spatial and temporal scales of the flux tubes ($\gtrsim$ few hundred km and $\sim$ few days). In fact, the numerical results are easily scalable to $2\,\mbox{km}$ with temporal scale $\sim 2$\,\mbox{s}. 
It should be pointed out that present observations can not resolve the shear flow scale since best achievable resolution is of Hinode is $\gtrsim 100—-200\,\mbox{km}$.  
 
As has been noted above, the spatial scale over which flow and field generation occurs is much smaller than the typical tube diameter. Thus we shall approximate the cylindrical tube by a planar sheet and work in the Cartesian coordinates where $x\,,y\,,z$ represent the local radial, azimuthal and vertical directions. We shall assume an initial homogeneous state with azimuthal shear flow $\v = {v_0}^{\prime}\,x\,\bmath{y}$ and uniform vertical field, $\bmath{B} = B\,\bmath{z}$. We note that collisional dissipation will cause the loss of energy in the lower solar atmosphere and in general a proper energy equation should be used for a more realistic modelling of the physical processes \citep{K11}. However, in order to keep the description simple, and to understand the underlying physical processes we shall assume that the fluid is incompressible. 

We shall assume that the solutions of the linearised equations is of the form $\exp{\left(\sigma\,t + i\,k\, z\right)}$. This simple form allows us to split the linear system of equations in two subsystems: one subsystem corresponding to sound waves propagating along the magnetic field and the interesting subsystem for which $z-\mbox{components}$ of the perturbed velocity, magnetic field, current density and electric field all vanish. 

The $x$ and $y$ components of the momentum equation yield
\bq
\frac{\dv}{v_A}
 = \frac{i\,\bk}{\bs^2}\,
\left(
\begin{array}{cc} \bs   &  0\\
                   \alpha  & \bs
  \end{array}
\right)\,\frac{\dB}{B}\,.
\label{eq:Lm}
\eq
 The magnetic drift velocity, Eq.~(\ref{eq:md0})  becomes
\bq
\frac{\dv_B}{v_A} = i\,\bk\,\left(
\begin{array}{cc} \betaP   &  \betaH\\
                   - \betaH & \betaP
  \end{array}
\right) \,\frac{\dB}{B}\,.
\label{eq:Ldv}
\eq
Since $\delta \Jpa = 0$, $x$ and $y$ components of the induction Eq.~(\ref{eq:indA}) can be written as   
\bq
\left(\begin{array}{cc} \bs   &  0\\
                   \alpha & \bs
  \end{array}
\right)\,\frac{\dB}{B} = i\,\bk\,\left(\frac{\dv}{v_A}
 + \frac{\dv_B}{v_A}
\right)\,.
\label{eq:Ld}
\eq
In the above Eqns.~(\ref{eq:Lm})--(\ref{eq:Ld}), wavenumber $k$, frequency $\sigma$ and diffusivity $\eta$ are normalised:  $\bk = k\,v_A / |\vp|$, $\bs = \sigma /|\vp|$ and $\betaP = \eta_p\,|\vp| / v_A^2$, $\betaH = \eta_H\,|\vp| / v_A^2$ . Here $v_A = B / \sqrt{4\,\pi\,\rho}$ is \alf speed and $\alpha = - \vp / \ |\vp| \equiv \pm 1 $. 
Making use of Eqs.~(\ref{eq:Lm})–-(\ref{eq:Ldv}), Eq.~(\ref{eq:Ld}) can be written as
\begin{eqnarray}
\left\{
\left(\begin{array}{cc} \bs   &  0\\
                   \alpha & \bs
  \end{array}
\right) 
+ \frac{\bk^2}{\bs^2}  
\left(\begin{array}{cc} \bs   &  0\\
                   \alpha & \bs
  \end{array}
\right) 
\right.
\nonumber\\
+
\left.
\bk^2\,\left(
\begin{array}{cc} \betaP   &  \betaH\\
                   - \betaH & \betaP
  \end{array}
\right)\right\}
\,
\frac{\dB}{B} = 0\,.
\label{eq:Ld1}
\end{eqnarray}
Since the dynamical time of interest here is set by shear flow gradient i.e. $\sigma \sim |\vp|$ or, $\bs \sim 1$ three terms in the preceding equation is of the order of $1\,,\bk^2\,,\bk^2\,\etas$, where $\etas = \sqrt{\betaH^2 + \betaP^2}$. Thus, Eq.~(\ref{eq:Ld1}) can be analysed in three different limits \citep{WS11}. 

I. {\it Ideal MHD: \footnote{Here ideal MHD of weakly ionized plasma pertains to the neutral fluid which carries the inertia of the medium.} Magnetic diffusion is negligible in comparison with the fluid advection, i.e. the field is frozen in the fluid; $\left[\dv / v_A \sim  \dB / B \right]$}. In this limit $\bk^2 \sim 1 \gg \bk^2\,\etas$, i.e. $\etas \ll 1$ and the last term in Eq.~(\ref{eq:Ld1}) can be neglected. Since $\bk^2 \sim 1$ implies $k\,v_A \sim |\vp|$ for $v_A = 5\,\mbox{km}/\mbox{s}$ and vortex flow $|\vp| \sim 0.1 /\mbox{s}$ \citep{SN98}, this gives $\lambda \sim 300\,\mbox{km}$. The pressure scale height $H$ in the photosphere is $\sim 150\,\mbox{km}$ \citep{S76}. Thus $\lambda \gg H$ and the ideal MHD limit is not applicable to the thin flux tubes. 

II. {\it Cyclotron limit: Magnetic diffusion is comparable to the fluid advection;}  $\left[\dv / v_A \gg  \dB / B\right]$. In this case field diffusion balances fluid convection, i.e. $\bk^2\,\etas \sim \bk^2 \gg 1$. This is the low frequency limit since $\bs \sim 1$ and first term in Eq.~(\ref{eq:Ld1}) can be neglected. The low frequency, short wavelength dressed ion—-cyclotron wave with frequency $\omega_{C} = \omega_H$ is the normal mode of the system \citep{PW08}. 

The requirement $ \bk^2\,\etas \sim \bk^2 \gg 1$ is never fulfilled in the solar atmosphere since $\etas \equiv \eta_{\perp}\,|\vp| / v_A^2 \sim 10^{-2}-10^{-4} \ll 1$. Therefore, this limit is not valid in the solar atmosphere.

III. {\it Highly diffusive limit: Magnetic diffusion overwhelms fluid advection;} $\left[\dv / v_A \ll \dB / B\right]$. In this case $\bk^2\,\etas \sim 1 \gg \bk^2$, i.e. $\etas \gg 1$ and $\bk^2 \ll 1$. The evolution of magnetic field which is kinematic in nature in this case is solely determined by diffusion. Only the first and last terms in Eq.~(\ref{eq:Ld1}) are retained.  For typical values of $\eta_{\perp}$ (Table~1), this limit gives $\lambda \lesssim 6\,\mbox{km}$ which fits within the pressure scale height. Therefore, we shall work in the high frequency limit and neglect the advection term in Eq.~(\ref{eq:Ld1}). 

In the highly diffusive limit, we get the following dispersion relation  
\bq
\sigma^2 + C_1\,\sigma + C_0 = 0\,,
\eq
where
\bq
C_1 =  2\,k^2\,\eta_P\,,\quad C_0 = k^2\,\left[k^2\,\eta_{\perp}^2 + \vp\,\eta_H\right]\,.
\eq
This dispersion relation can as well be written in non-dimensional form $
\bs^2 + C_1\,\bs + C_0 = 0\,,$ with coefficients 
\bq
C_1 = 2\,\bk^2\,\betaP\,,\quad C_0 = \bk^2\,\left(\bk^2\,\etas^2 - \alpha\,\betaH\right)\,.
\label{eq:mdr1}
\eq
In the absence of shear flow and $\eta_P$, the right circularly polarised whistler  
\bq
\omega_W =(k\,v_A)^2 / \omega_H \equiv k^2\,\eta_H\,,
\label{eq:whw}
\eq
is the normal mode of the system \citep{PW08}. 

The necessary condition for Hall instability, $C_0 < 0$, becomes
\bq
- \vp\,\eta_H > k^2\,\eta_{\perp}^2\,.
\label{eq:INc}
\eq
Note that the sign of Hall diffusion ($\eta_H$) depends on the orientation of the vertical magnetic field with respect to the shear flow gradient \citep{W99}.  As a result onset of the Hall instability depends not only on the sign of the velocity gradient but also on the orientation of the vertical magnetic field. Thus when the shear gradient is such that $- \vp > 0$, a positive $\eta_H$ (which can be guaranteed if the magnetic field is parallel to the shear gradient, $- \vp$) allows the above condition to be easily satisfied. When the magnetic field is anti—parallel to the shear gradient $\vp > 0$, above condition can as well be satisfied. Therefore, the orientation of the magnetic field together with the shear gradient plays crucial role in the Hall instability. 
This instability is the planar version of well known Hall-modified magnetorotaional instability in accretion discs where the differential rotation of the disc [$v(x) = x\,\Omega(x)$ where $\Omega$ is the orbital frequency] causes shear in the disc \citep{W99, WS11}. For definiteness in the subsequent analysis we shall assume $\alpha = 1$.  

Eq.~(\ref{eq:INc}) suggests that the fluctuations of wavelength
\bq
\lambda > \frac{2\,\pi}{\sqrt{|-\vp|}}\,\left(\eta_H + \frac{\eta_P^2}{\eta_H}\right)^{1/2}\,
\eq 
will become unstable. Flow gradients with either sign will give real $\lambda$ provided magnetic field has the right orientation.  The wavelength $\gtrsim 60\,\mbox{km}$ becomes Hall unstable for $\eta_H \sim 10^{10}\,\mbox{cm}^2 / \mbox{s}$ and $\eta_P \sim \eta_H$ (Table~1).  

The growth rate of Hall instability is
\bq
\bs = \sqrt{\bk^2\,\betaH\, \left(1 -– \bk^2\,\betaH\right)} –- \bk^2\,\betaP\,. 
\label{eq:grw}
\eq

\begin{figure}
\includegraphics[scale=0.30]{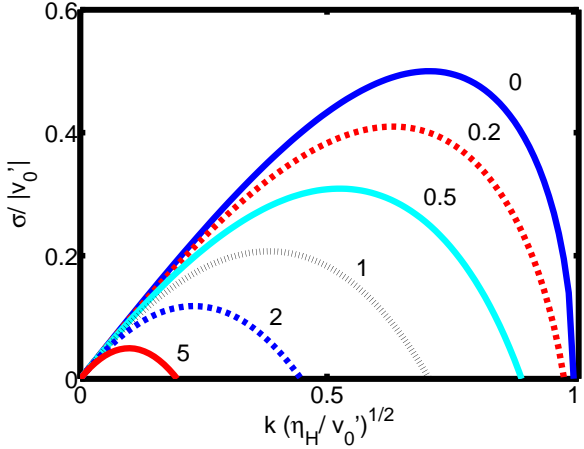}
\caption{The growth rate of Hall instability $\bs \equiv \sigma / |\vp|$ against $\bk\,\left(\betaH \right)^{1/2} \equiv k\,\left(\eta_H / |\vp|\right)^{1/2} $ is shown for different values of $\eta_P / \eta_H$.}
 \label{fig:FT}  
\end{figure}

In Fig.~ (\ref{fig:FT}) Eq.~(\ref{eq:grw}) is plotted for various values of $\eta_P / \eta_H$ and $\alpha = 1$. We see that with increasing $\eta_P$ the growth rate decreases before completely disappearing when $\eta_P \gg \eta_H$, that is when the available free energy is completely dissipated by Pedersen diffusion. 

The physical picture of Hall instability is quite simple. The shear flow generates $\dBy$ from $\dBx$ and Hall diffusion generates $\dBx$ from $\dBy$. 
This is how Hall diffusion in tandem with shear flow destabilises the flux tube. This also explains the dependence of growth rate on both shear gradient and whistler frequencies. 

We can find the maximum growth rate of Hall instability by recasting the dispersion relation, Eq.~(\ref{eq:mdr1}) in $a\,k^4 + b\,k^2 + c = 0$ form and setting the discriminant $b^2 -– 4\,a\,c = 0$. Here  
  \begin{eqnarray}
a = \etas^2\,,\,\,b = 2\,\betaP\,\bs - \betaH\,,\,\,c = \bs^2\,.
\label{eq:coeff2}
\end{eqnarray}
The maximum growth rate of the instability becomes 
\bq
\sigma_0 = \left(\frac{\eta_H}{\eta_P + \eta_{\perp}}\right)\,\frac{|\vp|}{2} 
\equiv \left(\frac{1}{2}\right)\,\frac{|\vp|}{\eta_P / \eta_H + \sqrt{1 + \eta_P^2 / \eta_H^2}}\,.
\label{eq:hmax1}
\eq
It is clear from the preceding equation that the maximum growth rate is proportional to the shear gradient and to leading order, is inversely proportional to $1 + \eta_p / \eta_H$. Thus, in conformity with Fig.~ (\ref{fig:FT}),  the maximum growth rate of the instability corresponds to $\eta_P / \eta_H = 0$ and with increasing $\eta_p / \eta_H$, the instability grows at smaller rates.  The wavenumber corresponding to maximum growth rate is 
\bq
k_0 = \sqrt{\frac{\eta_H\,|\vp|}{2\,\eta_{\perp}\left(\eta_P + \eta_{\perp}\right)}}\,.
\label{eq:hk}
\eq
Thus the maximum growth rate is $\sigma_0 = k_0^2\,\eta_{\perp}$. Note that $k_0 \propto 1 / \sqrt{\eta_H} \sim 1 / \sqrt{B}$ and hence for a given $\eta_P / \eta_H$, a weaker field implies that smaller wavelengths are Hall unstable.     

\section{Application to the solar atmosphere}
Intense magnetic fields are organised as mainly vertical flux tubes and are thought to attain steady state possibly due to the pressure balance with the ambient medium. Cylindrical coordinates are best suited to describe these structures. However, in order to develop proper understanding of physics behind the Hall diffusion driven shear flow instability, we have approximated cylindrical flux tubes locally by planar sheets in the present work where $x$ and $y$ coordinate locally correspond to the radial and azimuthal directions respectively on a cylinder. 

For a kiloGauss field, Hall diffusion dominates photosphere and lower chromosphere between $0$ and $1\,\mbox{Mm}\,$ (see Fig.~1).  Thus, in active and quiet solar regions closer to the footpoint ($h = 0$) where kG field may be present, Hall instability will have the maximum growth rate which is on the order of the shear gradient [$\eta_P = 0$ in Fig.~ (\ref{fig:FT})]. With increasing height, for example at $h \sim 1.5\,\mbox{Mm}$, when $\eta_P / \eta_H \sim 4$ (Fig.~1), the Hall instability growth rate is one third of shear gradient. At $h \sim 2\, \mbox{Mm}$, when $\eta_P / \eta_H = 1.75$ the growth rate reduces to $\sim 0.35\,|\vp|$. Therefore, with increasing height ($\gtrsim 1\,\mbox{Mm}$), when Pedersen diffusion dominates Hall diffusion, Hall instability grows at considerably smaller rate than below $1\,\mbox{Mm}$. 

In the internetwork regions, when $B_0 \sim 100\,\mbox{Gauss}$, Hall diffusion dominates Pedersen in the entire photosphere-–chromosphere except for a very small region closer to the solar surface ($\lesssim 0.25\,\mbox{Mm}$) (dotted curve in Fig.~1). As a result, Hall instability will grow at a rate close to shear frequency (curves labelled $0$ and $0.2$ in Fig.~ [\ref{fig:FT}]), in the entire photosphere--chromosphere above $0.25\,\mbox{Mm}$.

The maximum growth rate of Hall instability is independent of the magnetic field strength only when $\eta_P / \eta_H \ll 1$. This implies that for the kG network  or internetwork field below $1\,\mbox{Mm}$ or, weak internetwork fields above $0.25\,\mbox{Mm}$ (where Hall is the dominant diffusion), Hall instability will not be affected by decreasing magnitude of the field. However, in the weak—-field internetwork regions below $\lesssim 0.25\,\mbox{Mm}$, where $\eta_P / \eta_H \gtrsim 1$ the maximum growth rate can be quenched by the presence of a strong field, since $\eta_P \sim B^2$ and $\eta_H \sim B$, and thus the maximum growth rate is inversely proportional to the magnetic field. Therefore, in strong field regions, long wavelength fluctuations are suppressed by Pedersen diffusion.  

In the present problem we have assumed that the partially ionized plasma is threaded by a uniform vertical magnetic field. This is the simplest representation of magnetic flux tubes in the solar atmosphere. We could directly apply these results to flux tubes provided the flow velocity and shear scale were known. Therefore, we turn to recent observation and numerical simulations. The vortex flows on the solar surface have been observed by various groups [Please see excellent review by Steiner \& Rezaei (2012)]. Small whirlpools similar to terrestrial hurricanes ($\lesssim 500\,\mbox{km}$) with mean lifetime ($\sim 5\,\mbox{min.}$) were discovered by \cite{B08}. Larger vortices with longer lifetime and possible entanglement of magnetic field manifested as enhanced CaII emission at vortex centre have also been observed \citep{A09}. The typical vorticity of a vortex is $\sim 6\times 10^{-3}\,\persec$ which corresponds to rotation period $\sim 35\,$ minutes \citep{B10}. Thus it would appear that Hall instability does not have time to develop since the growth rate ($\avp / 2 = 3\times 10^{-3}\,\persec$) is very small. However, above vorticity value is limited by the upper limit in the vorticity resolution ($\sim 4\times 10^{-2}\,\persec$, \cite{B10}). The numerical simulation gives much higher vorticity value ($\sim 0.1—-0.2\,\persec$) in the photosphere—-lower chromosphere (Fig.~31, \cite{SN98}). The growth rate corresponding to $\avp = 0.2\,\persec$ is one minute.  

For a kilo Gauss field at the footpoint of the flux tube ($h = 0$), using corresponding number densities of neutrals and ions (model C, VAL81), the \alf speed becomes $v_A \sim 5 \times 10^5\,\mbox{cm} / \mbox{s}$. In Table~1, we give values of Hall diffusion the photosphere-–chromosphere region which has been used to estimate fluctuation wavelengths. Since $\eta_H \sim B$, the value of Hall diffusivity in the internetwork region with weak field $\sim 120\,\mbox{G}$ will be an order of magnitude smaller than values given in the table.  
\begin{table*}
 \centering
 \begin{minipage}{120mm}
  \caption{\label{tab:table1} The values of Hall diffusivity at different altitude are given in the table for a magnetic field profile $B = B_0 \left( n_n / n_0\right)^{0.3}$ with $B_0 = 1.2\,\mbox{kG}$ and $n_0 = 1.16\cdot10^{17}\,\mbox{cm}^{-3}$ at $\mbox{h} = 0$ \citep{VAL81}.}
    \begin{tabular}{|@{}llrrrrrrlll|@{}|}
    \toprule[0.12em]
h\,(km) & $0$  & $250$ &$515$ &  $1065$& $1515$& $2050$&$2298$&$2543$\\
\midrule[0.12em]
 $\eta_H [10^7\,\mbox{cm}^{2} / \mbox{s}] $ & $9.3$ & $130$ & $710$& $460$ & $290$ & $220$ &$960$&$2100$\\
\bottomrule[0.12em]
\end{tabular}
\end{minipage}
\end{table*}

\begin{figure}
\includegraphics[scale=0.30]{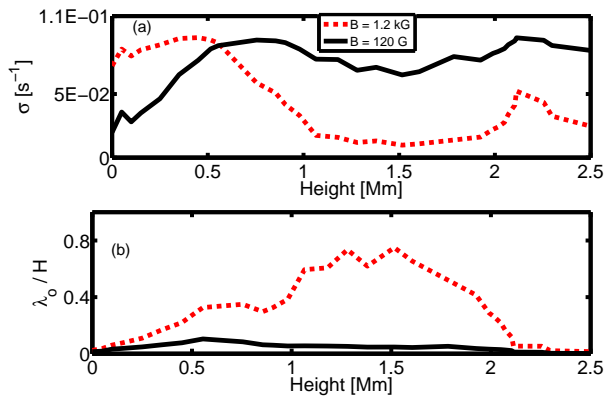}
\caption{The maximum growth rate and most unstable wavenumber normalised to pressure scale height for $1.2\,\mbox{kG}$ (bold line) and $120\,\mbox{G}$ field (dotted line) against altitude measured from solar surface ($ h = 0$) is shown in Fig.~4(a) and Fig.~4(b) respectively.}
 \label{fig:FT1}  
\end{figure}
In Fig.~[\ref{fig:FT1}(a)--(b)], we plot the maximum growth rate, Eq.~(\ref{eq:hmax1}) and corresponding wavelength Eq.~(\ref{eq:hk}) for $|\vp| = 0.1\,\mbox{s}^{-1}$ against altitude. As expected the maximum growth rate ($\sim |\vp|$) for kilo Gauss field occurs around $0.5\,\mbox{Mm}$, where $\eta_P / \eta_H \ll 1$ (Fig.~\ref{fig:DF}). With increasing altitude, since Pedersen becomes comparable to Hall diffusion, the growth rate tapers off and beyond $\sim 1\,\mbox{Mm}$ Hall instability may not be dynamically important. However, for a $120\, \mbox{G}$ field, excluding a small region in the lower photosphere, the maximum growth rate remains robust, $\lesssim |\vp|$ in entire photosphere-–chromosphere. 

In Fig.~\ref{fig:FT1}(b) we plot the most unstable wavelength normalised against pressure scale height $H$ which has been calculated using model C, VAL81. For both intense and weak fields,  $\lambda_0 \lesssim H$ and the fluctuation wavelength fits within the scale height. Therefore, the most unstable wavelength fits within the scale height in the entire photosphere-—chromosphere. 

It would appear from Eq.~(\ref{eq:mdr1}) that the wavenumber spectrum has finite cut—off. However, cut—-off wavelength can not be inferred from Eq.~(\ref{eq:mdr1}) as $\bs \sim 1 $ and $\bk \sim 1 / \sqrt{\tilde{\eta}}$ in highly diffusive limit. In the $\bs \rightarrow 0$ limit, the neglected advection term in the induction equation also becomes important. Therefore, we can not infer cut—-off wavelength in the highly diffusive limit. The spectrum of Hall instability depends on the ambient magnetic field strength, i.e. $k_0 \propto 1/ \sqrt{B}$ implying that the wave spectrum will have sharper peak in kG field regions.  

With increasing altitude vertical tubes expand and bend developing small azimuthal field components in the process. As will be shown elsewhere \citep{PW12}, growth of the Hall instability is the merely rescaled due to bending of the field. 

\section{Discussion and summary}
Granular motions are responsible for the generation of low frequency waves in the predominantly neutral photosphere and lower chromosphere. In the presence of a shear flow gradient, tubes in both network and internetwork regions can become unstable due to Hall diffusion. Recall that the instability growth rate is $\vp / 2$ which for $\vp = 0.1 \,\mbox{s}^{-1}$ correspond to the e-fold time $\sim 20\,\mbox{s}$. Since average period of vortical motion in the photosphere is $\sim 300\,\mbox{s}$, Hall instability have sufficient time to develop in the vortices. The observed value of vorticity is an order of magnitude smaller \citep{B10} than numerical value suggesting that Hall instability may not have enough time to develop. However, observed small value of the vorticity is due to limited spatial—-temporal resolution \citep{B10}. The simulation results indicate that highest vorticities occur at the smallest observable scale \citep{SN98}. Therefore, we conclude that Hall instability in tandem with shear flow may destabilise the flux tubes. 

Can Hall instability excite turbulence and heat the medium?  This question falls beyond the scope of present linear analysis. Only numerical simulation can provide an answer to this question. Numerical simulations are approaching length scales associated with the required shear scale \citep{M11b, C12, SK12}. Therefore, we hope that simulation by various groups will be able to delineate the role of Hall instability in chromospheric heating. The present analysis provides plausible pathway to such a process. 

\section*{Acknowledgments}
{The financial support of Australian Research Council through grant DP0881066 is gratefully acknowledged. This research has made use of NASAs Astrophysics Data System.}

\end{document}